\documentclass[a4paper,10pt,fleqn,usenames,dvipsnames]{article}

\usepackage{lmodern}
\usepackage[T1]{fontenc}
\usepackage{amsmath}
\usepackage{amssymb}

\usepackage{url}
\usepackage{longtable}
\usepackage{multirow}
\usepackage{caption}

\usepackage[nodayofweek]{datetime}
\usepackage{enumerate}
\usepackage{xspace}
\usepackage{xcolor}
\usepackage{graphicx}
\usepackage[pdftex,hidelinks]{hyperref}
\usepackage{bookmark}
\usepackage{pgffor}
\usepackage{subfig}
\usepackage{float}
\usepackage[capitalise]{cleveref}

\usepackage[backend=biber,giveninits=true,doi=false,sorting=custom,style=numeric-comp,url=false,isbn=false,eprint=false]{biblatex}
\DeclareFieldFormat[article]{citetitle}{\textit{#1}\isdot}
\DeclareFieldFormat[article]{title}{\textit{#1}\isdot}
\DeclareFieldFormat[unpublished]{citetitle}{\textit{#1}\isdot}
\DeclareFieldFormat[unpublished]{title}{\textit{#1}\isdot}
\DeclareFieldFormat[inproceedings]{citetitle}{\textit{#1}\isdot}
\DeclareFieldFormat[inproceedings]{title}{\textit{#1}\isdot}

\DeclareNameAlias{author}{last-first}

\AtEveryBibitem{\clearfield{pages}} 
\AtEveryBibitem{\clearfield{volume}} 
\AtEveryBibitem{\clearfield{month}} 
\AtEveryBibitem{\clearfield{day}} 
\AtEveryBibitem{\clearfield{note}} 
\AtEveryBibitem{\clearfield{addendum}} 
\AtEveryBibitem{\clearfield{Addendum}} 
\AtEveryBibitem{\clearfield{type}} 
\AtEveryBibitem{\clearfield{address}} 
\AtEveryBibitem{\clearfield{institution}} 

\newbibmacro{string+doi}[1]{%
  \iffieldundef{doi}{\iffieldundef{url}{#1}{\href{\thefield{url}}{#1}}}{\href{http://dx.doi.org/\thefield{doi}}{#1}}}	
\DeclareFieldFormat*{title}{\usebibmacro{string+doi}{\mkbibemph{#1}}}

\DeclareSortingScheme{custom}{
    \sort{
        \citeorder
    }
    \sort{
        \field{year}
    }
}

\usepackage{xstring}
\newcommand{\Cwola}{\textsc{CWoLa}\xspace}
\newcommand{\scwola}{strong \textsc{CWoLa}\xspace}

\newcommand*{\TeV}{\,\text{TeV}}
\newcommand*{\GeV}{\,\text{GeV}}

\usepackage[nonatbib,preprint]{neurips}

\title{Strong \Cwola: Binary Classification Without Background Simulation}
\author{%
  Samuel Klein \\
  University of Geneva\\
  \texttt{samuel.klein@unige.ch} \\
  \And
  Matthew Leigh \\
  University of Geneva\\
  \texttt{matthew.leigh@unige.ch} \\
  \And
  Stephen Mulligan \\
  University of Geneva\\
  \texttt{stephen.mulligan@unige.ch} \\
  \And
  Tobias Golling \\
  University of Geneva\\
  \texttt{tobias.golling@unige.ch} \\
}

\begin{document}
    \maketitle
    
    \begin{abstract}
      Supervised deep learning methods have been successful in the field of high energy physics, and the trend within the field is to move away from high level reconstructed variables to lower level, higher dimensional features. 
      Supervised methods require labelled data, which is typically provided by a simulator. 
      As the number of features increases, simulation accuracy decreases, leading to greater domain shift between training and testing data when using lower-level features.
      This work demonstrates that the classification without labels paradigm can be used to remove the need for background simulation when training supervised classifiers. 
      This can result in classifiers with higher performance on real data than those trained on simulated data.
    \end{abstract}
    

    \section{Introduction}
    Supervised machine learning (ML) techniques have revolutionized data analyses in the field of high energy physics (HEP)~\cite{nature_hep_ml} with huge advancements in flavor~\cite{FTAG-2019-07,ATL-PHYS-PUB-2023-021} and jet tagging~\cite{Kogler_2019,Larkoski_2020}. 
Part of the success of these methods is due to architectural improvements, such as the use of transformers~\cite{Mikuni_2021,qu2024particletransformerjettagging} and graph neural networks~\cite{pelican,Moreno_2020,dreyer2021jettagginglundplane,Shlomi_2021,Gong_2022}, but it is also due to the increase in the size of datasets and a shift away from fitting models on high level reconstructed variables towards low level variables that are closer to what is directly measured by experiments.
These low level features contain more information and therefore allow for the development of more powerful classifiers.

High fidelity simulators are used to generate labelled training data for supervised ML models~\cite{Buckley_2011,ATLAS:2010arf,AGOSTINELLI2003250}.
However, simulation is not perfect due to a variety of factors, such as the complexity of the underlying physics processes and unknowns in the detector response~\cite{ATLAS:2024rua}. 
Due to these imperfections, there is a distribution shift between simulated and real data.
This shift means that models trained purely on simulation underperform on real data~\cite{muonsdata} and become harder to use with significant calibration effects~\cite{Andreassen_2020,Pollard_2022,flowaway}.
The shift is larger on lower level data as the information content increases.
Mitigating these effects is a difficult but important problem for the field.

A task of particular interest in HEP is the binary classification task of discriminating signal from background.
Models that target a specific signal can be trained using data only by assigning weak labels and exploiting the classification without labels (\Cwola) paradigm~\cite{Metodiev_2017}.
For muon isolation this approach has been shown to improve over training on simulation~\cite{muonsdata} due to the absence of simulation mismodelling when using real data.
However, applying the standard \Cwola approach to a supervised task requires defining two measured datasets with different signal-to-background mixtures, where each component in both datasets is sampled from the same data generation processes.
While this was shown to be possible for muon isolation~\cite{muonsdata}, it is not possible to define such datasets for all signal models.
For this reason, the \Cwola approach has mostly been applied to searches where no signal model is assumed and assumptions are instead made about the background distribution rather than targeting a specific signal~\cite{Andreassen:2020nkr,Benkendorfer_2021,drapes,Golling:2022nkl,Cathode,Hallin_2023,curtainsf4f,leigh2024acceleratingtemplategenerationresonant,curtains,HDBS-2018-59}.
These weakly supervised searches are challenging because no strict assumptions about the background distribution can be made, and therefore the background modelling is often low quality.
Further, the standard \Cwola approach struggles when there is only a small total number of signal events in the dataset~\cite{Finke_2022}, and this is the setting that is most interesting for new physics searches.

This work demonstrates that \Cwola can be applied to any strongly supervised settings where simulated signal is available.
The method will be referred to as strong \Cwola (s\Cwola), and it allows supervised binary classifiers to be trained without the use of background simulation.
This work shows that \scwola matches or improves upon the performance of training on simulated background, and can be applied to any binary classification task where simulated signal can be produced.
The \scwola approach is applicable to searches for specific new physics processes.

    \section{Strong \Cwola}
    Consider two datasets $M_1$ and $M_2$ with different signal fractions, $f_1$ and $f_2$, respectively.
If $f_1 > f_2$ and both signal and background are sampled from the same distribution in both datasets, then the optimal classifier for discriminating signal from background can be found by learning to discriminate $M_1$ from $M_2$.
This defines the \Cwola paradigm and forms the basis of the \scwola method.

\begin{figure}
    \centering
    \includegraphics[width=0.5\textwidth]{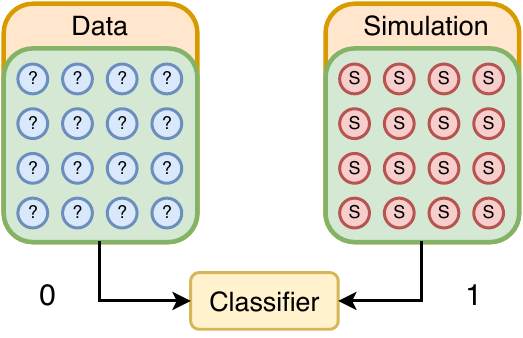}
    \caption{Outline of the \scwola method. The data is labelled zero and identified with a question mark to highlight that the nature of these samples is not known, and the dataset contains a possibly unknown mixture of signal and background. The signal samples are generated with a simulator and are all labelled one as the origin of these events is known exactly.}
    \label{fig:strong_cwola_outline}
\end{figure}

In the \scwola approach the dataset $M_1$ is composed of simulated samples of the targeted signal process; this dataset is pure signal ($f_1=1$).
The dataset $M_2$ is real data, an unknown mixture of signal and background.
All HEP datasets contain some background samples, but they may be devoid of signal, therefore $f_2 \in [0, 1)$.
This construction guarantees that $f_1 > f_2$ and the optimal classifier for discriminating between signal and background can be found by training a classifier to distinguish between pure simulated signal and data.
An outline of the \scwola algorithm is presented in \Cref{fig:strong_cwola_outline}.
The benefit of \scwola is that a classifier can be trained without any background mismodelling.
A classifier trained in this way will enhance signal in data, rather than the traditional approach of enhancing signal in simulated background.

    \section{Resonant search}
    This section demonstrates the utility of the \scwola method in the context of a search for resonant new physics.
The dataset used for this study was produced for the LHC Olympics 2020 community challenge~\cite{lhco4536377}.
The background samples are QCD dijet events with $p_T = 1.3 \TeV$ and the signal samples are generated from $W' \rightarrow XY$ with $m_{W'}=3.5\TeV$ and particles $X,Y$ with masses $500$ and $100 \GeV$.
Simulations of the signal and background were performed using Pythia 8~\cite{SJOSTRAND2008852,Sj_strand_2006}, while Herwig++~\cite{B_hr_2008} is used to simulate an alternative background sample.
Delphes 3.4.1~\cite{delphes,Mertens_2015,Selvaggi_2014} is employed for the detector simulation with the CMS detector card. 
Jet clustering was conducted on particle flow objects using the Fastjet software~\cite{Cacciari_2012,Cacciari_2006} with the anti-kt algorithm~\cite{Cacciari_2008} and a jet radius parameter of one.

In this work, classifiers are trained to discriminate between datasets labelled one and zero.
Pure simulated signal samples are always labelled one.
The Herwig dataset is treated as pure simulated background and always labelled zero.
The Herwig dataset is used to train models that are proxies for the traditional approach where models are trained on simulated background. 
Pythia datasets of both pure background and mixtures of signal and background are explored.
These mixtures are treated as data and always labelled zero, different mixtures are used to probe the impact of signal contamination.
Mismodelling of the signal process is not accounted for in this work.

\begin{figure}[htbp]
    \centering
    \subfloat[]{
        \includegraphics[width=\textwidth]{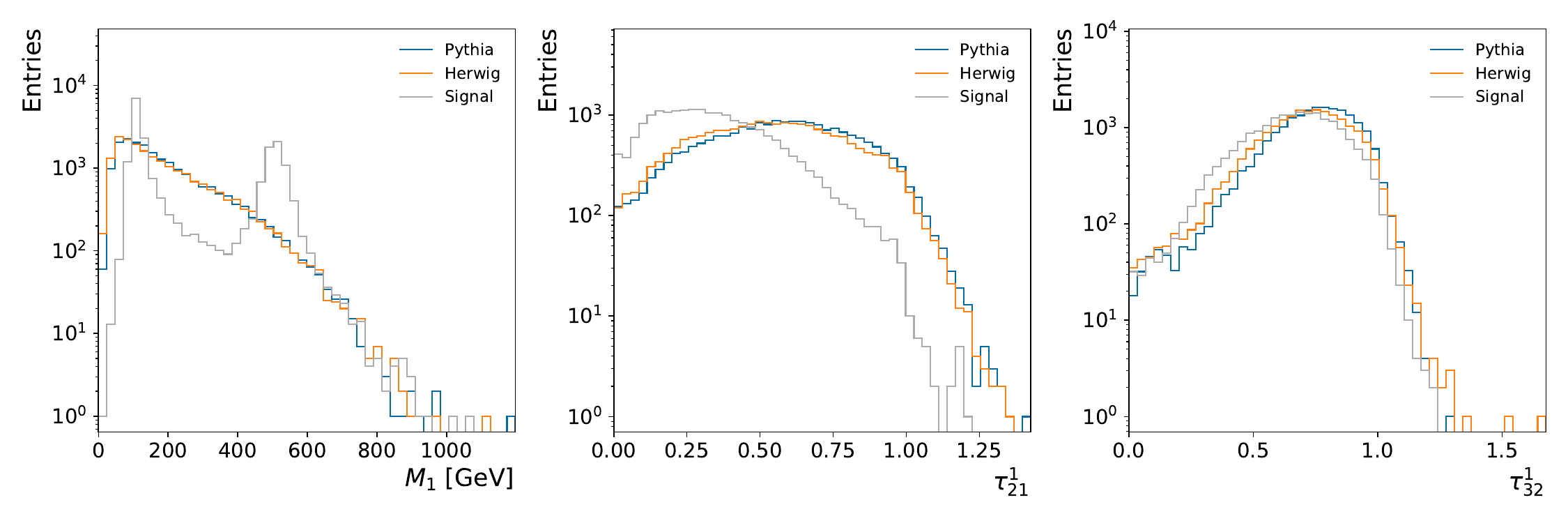}
    } \\
    \vspace{-3.5em}
    \subfloat[]{
        \includegraphics[width=\textwidth]{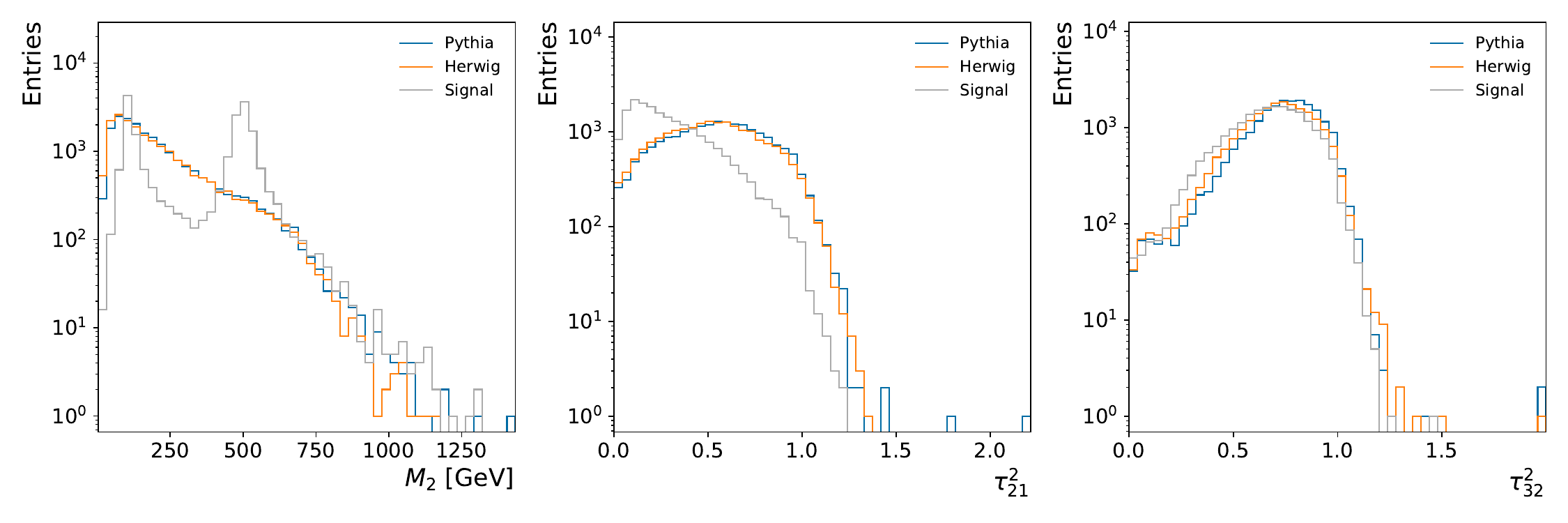}
    }
    \vspace{-2em}
    \caption{Distribution of high level features used for training. The Pythia and Herwig backgrounds are shown in blue and orange respectively, while the signal sample is shown in grey.}
    \label{fig:feature_hls}
\end{figure}

\begin{figure}
    \centering
    \includegraphics[width=\textwidth]{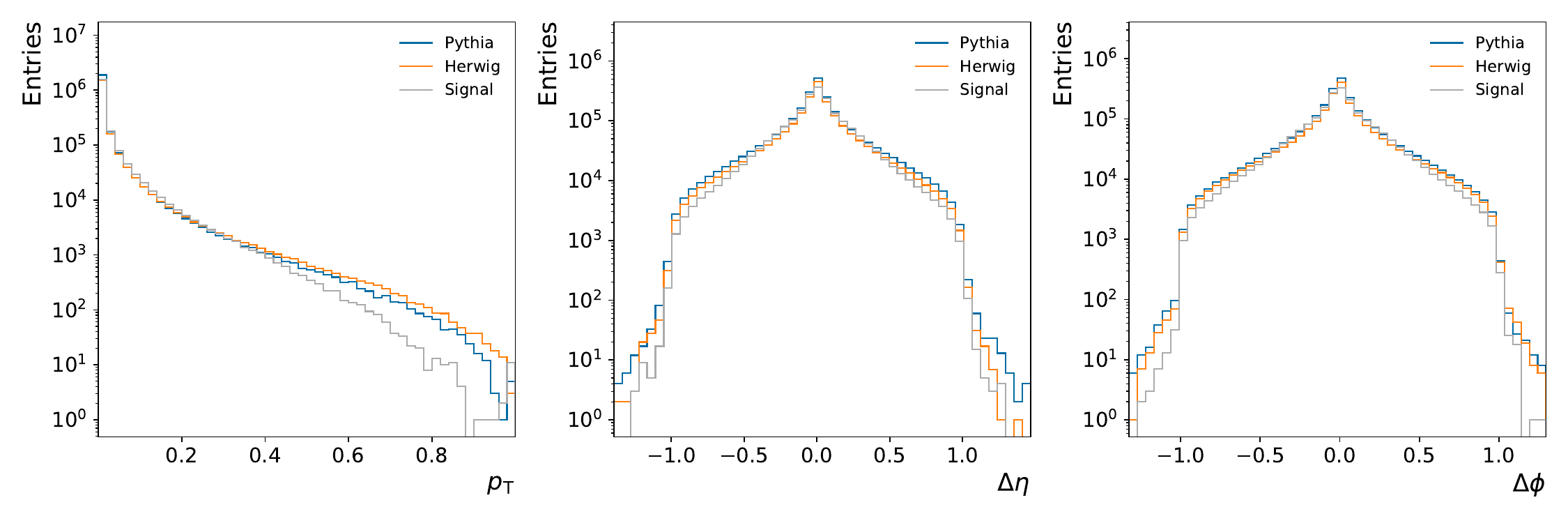}
    \vspace{-2em}
    \caption{Distribution of constituent features for both jets.}
    \label{fig:low_level_features}
\end{figure}

Both high and low level features are considered to demonstrate the impact of increased mismodelling.
For jet $i$ the mass $M_i$, 2- 1-subjettiness (3- 2-subjettiness) ratio $\tau_{21}^i$ ($\tau_{32}^i$) are used as high level features~\cite{PhysRevLett.105.092002}.
In the low level representation jets are an unordered point cloud, where each particle is described by its leading 100 constituents.
Each constituent is described by transverse momentum $p_{\mathrm{T}}$ and their coordinates in $\eta-\phi$ space relative to the central axis of the jet $\{\Delta \eta, \Delta \phi\}$.
The two background samples are similarly distributed over the high level features as shown in \Cref{fig:feature_hls}.
This is expected as our simulation is a close but imperfect representation of reality over these features.
The Pythia and Herwig samples are similarly distributed over the marginals of low level features as shown in \Cref{fig:low_level_features}.
It is expected that there will be more differences between these samples in higher order correlations.

To avoid training and evaluating the classifier on the same data, different folds are defined in a $k$-fold training strategy with $k=5$~\cite{hastie01statisticallearning_ch7, Aad_2020}.
Three fifths of the data are used to train each classifier, with one fifth for validation and the final fifth as the test set.
The full data set is cycled through such that each sample appears in the test set once.
For the high level variables Gradient Boosted Decision Trees (BDTs) implemented in \texttt{scikit-learn}~\cite{scikit-learn} are used as the classifier following the recipe defined in \textcite{Finke:2023ltw}.
For the low level variables a transformer model similar to that used in \textcite{drapes} is used.
This model is permutation invariant to the constituents of each jet and the order of the two jets.
For this, a transformer encoder with self-attention~\cite{Vaswani:2017lxt} is used on the jet constituents.
The low level classifier will also be pretrained using the masked particle modelling technique~\cite{leigh2024tokenizationneededmaskedparticle} and then fine-tuned on the binary classification task.
The same pretraining scheme is used for all low level classifiers.
For pretraining the model is trained on the combined Pythia and Herwig samples, including signal samples.
This pretraining is used as it has been shown to significantly improve the performance of \Cwola classifiers~\cite{leigh2024tokenizationneededmaskedparticle,golling2024maskedparticlemodelingsets}.
All details of the ML models and trainings are defined in \Cref{app:model_details}.

To explore the impact of signal contamination, the Pythia dataset is mixed with the signal dataset at different fractions.
Assuming a perfect background model for the QCD dijet mass background in 3.3-3.7 TeV, within which the signal is largely contained, a total signal contamination of ~1700 samples would result in a $5\sigma$ discovery.
Therefore, we explore the impact of signal contamination at and significantly above this boundary.
Demonstrating that the \scwola method is robust to signal contamination in this regime is important for the method to be useful in practice.
The \scwola method will be tested with 1700 and 5000 signal samples mixed with the Pythia background, this provides a test at the boundary of a $5\sigma$ discovery and a test at a higher signal contamination, respectively.

\begin{figure}[htbp]
    \centering
    \subfloat[]{
        \includegraphics[width=0.33\textwidth]{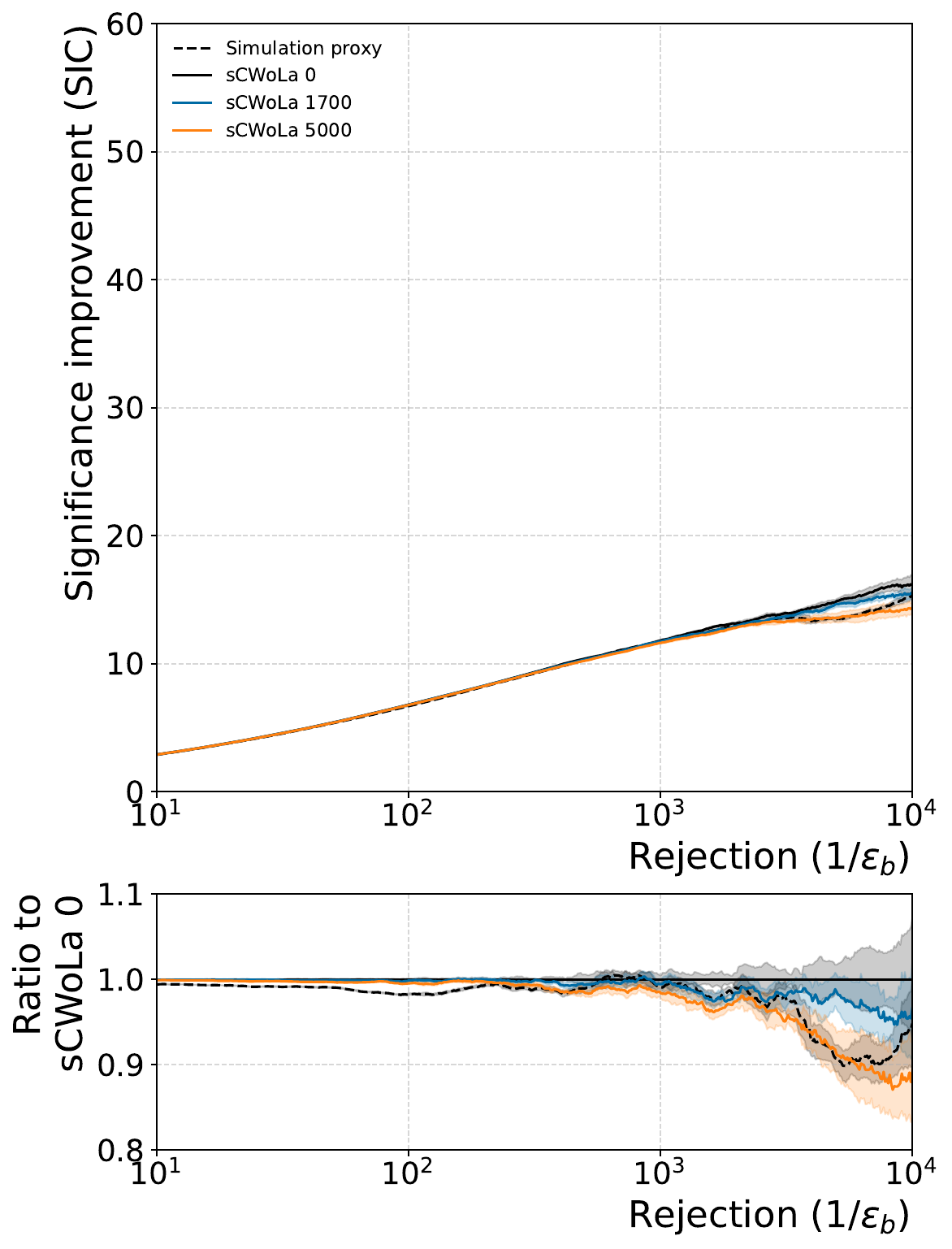}
 }
    \subfloat[]{
        \includegraphics[width=0.33\textwidth]{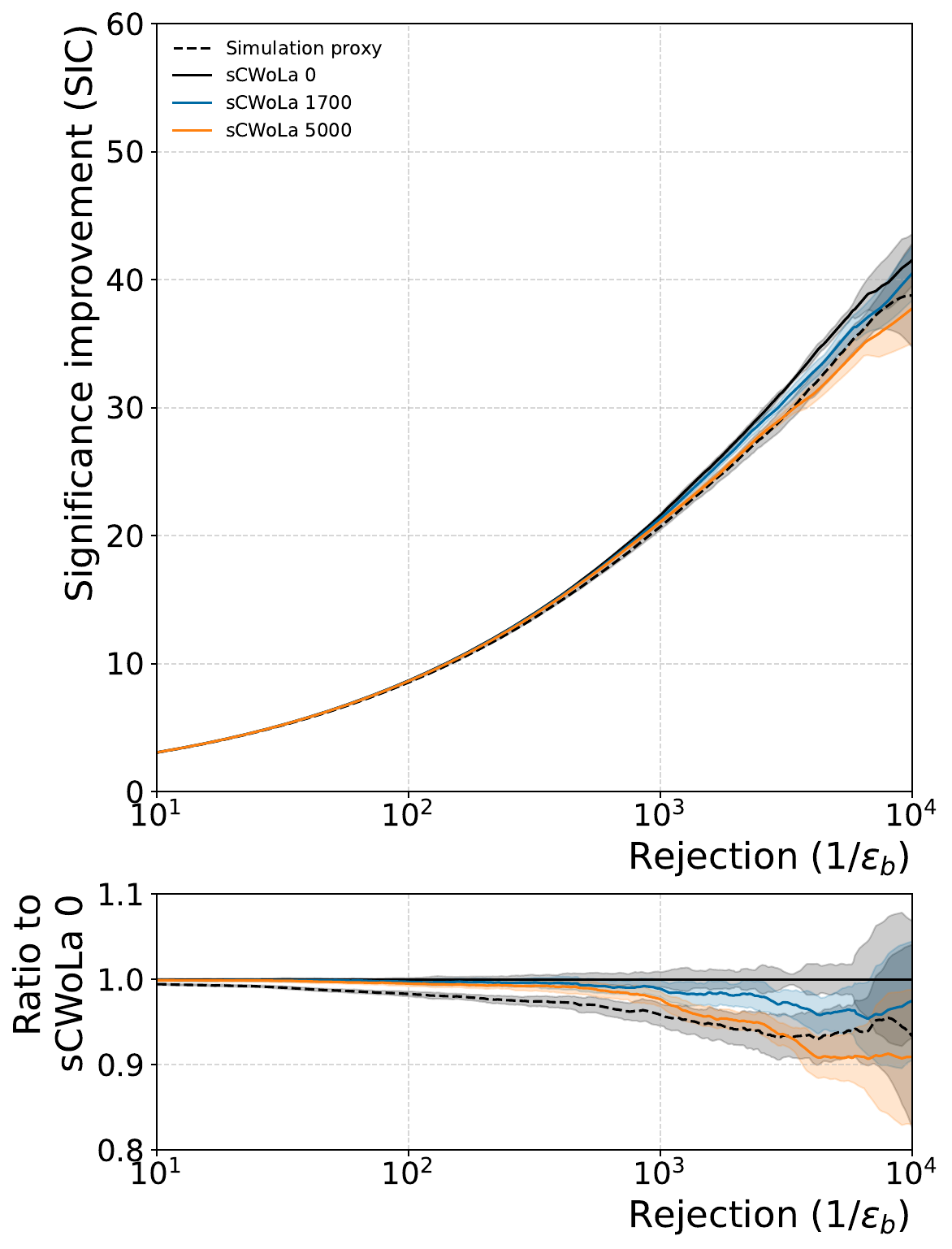}
 }
    \subfloat[]{
        \includegraphics[width=0.33\textwidth]{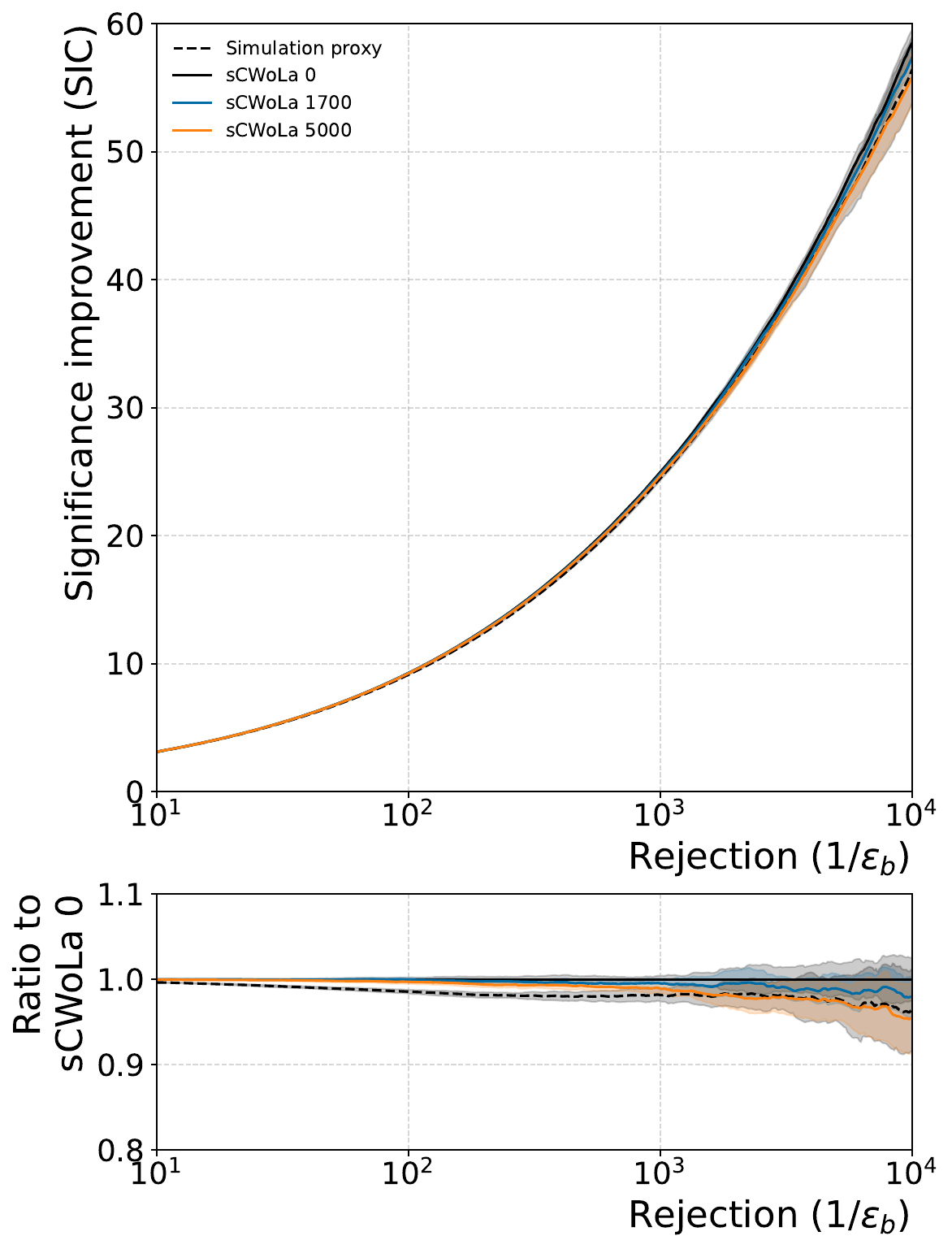}
 }
    \vspace{-2em}
    \caption{Significance improvement as a function of background rejection for the \scwola method (s\Cwola) at different levels of signal contamination and a classifier trained on Herwig background (Simulation proxy). The strong \Cwola classifier with no signal injected is the optimal classifier that can be achieved. A classifier trained on high level features (left), a classifier trained on low level features without pretraining (middle), and a pretrained classifier trained on low level features (right).}
    \label{fig:sic_hls}
\end{figure}

For all classifiers, the \scwola method can be seen to perform similar or better to the traditional approach as shown in \Cref{fig:sic_hls}.
The impact of signal contamination is seen to be small for both high and low level features.
There are some small differences at very high rejections where the finite statistics of the dataset is expected to have an impact.
The performance of the models increases significantly when using low level features. 
Pretraining the classifiers for use with low level features can be seen to improve the performance of all methods and largely removes any performance differences between the traditional and \scwola methods.
Training on a simulation proxy (Herwig) and evaluating on a data proxy (Pythia) -- the traditional classifier training approach -- is seen to perform close to the optimal setting of training on pure Pythia background against signal (s\Cwola 0).
In settings where the simulation is not a good representation of the data, the \scwola method is expected to outperform this traditional approach.
Further, even in the presence of perfect simulation, the ability to use the large amount of statistics present in many HEP analyses is expected to improve the performance of the \scwola method.

    \section{Conclusion}
    This work has presented the strong \Cwola approach, a new method for training supervised classifiers without using background simulation.
    The approach results in similar classifiers to the traditional approach and is applicable to any binary classification task where simulated signal can be produced and there is sufficient data in the targeted region of phase space to train a classifier.
    Classifiers can be trained on data using the $k$-fold approach to avoid training and evaluating on the same data.
    A robust evaluation within the context of a real analysis is needed to fully understand the potential of the strong \Cwola technique.
    In particular, taking into account systematic uncertainties on simulated signal will be crucial in the evaluation of the method~\cite{rothen2024enhancinggeneralizationhighenergy}.
    Further, while the strong \Cwola approach defines a way of making more optimal selections, a background estimate will still be needed in the context of a hypothesis test.
    Minimizing the impact of mismodelling in this estimate due to the classifier selection is an interesting avenue for future work.
    It will also be interesting to understand how the strong \Cwola approach benefits from the huge volumes of data available for analyses in HEP.

    The strong \Cwola method would be particularly useful for optimizing analyses in end-to-end frameworks, where the full analysis chain is optimized simultaneously.
    These approaches have been shown to improve the sensitivity of searches for new physics~\cite{lukas_finetuning}, but their performance can be expected to be limited by the quality of the simulation for low level features.
    The strong \Cwola approach would also complement foundation models trained on data~\cite{golling2024maskedparticlemodelingsets} particularly well, where supervised classifiers could be efficiently trained on huge volumes of data.

    The strong \Cwola approach has been presented within the context of HEP, but it could be applied to any binary classification task where direct access to samples from the signal distribution are available and there is sufficient measured data to train a classifier.
    Signal samples could be provided by a simulator, or human annotation, as long as signal samples are drawn from the correct target distribution the strong \Cwola approach will learn the optimal supervised classifer for isolating signal.

    \section*{Acknowledgements}
    The authors would like to acknowledge funding through the SNSF Sinergia grant CRSII5\_193716 ``Robust Deep Density Models for High-Energy Particle Physics and Solar Flare Analysis (RODEM)''
and the SNSF project grant 200020\_212127 ``At the two upgrade frontiers: machine learning and the ITk Pixel detector''.

    \appendix
    \section{Model details}
    \label{app:model_details}
    \subsection{High level classifier}
The classifier that uses high level variables is a Gradient Boosted Decision Trees implemented in the \texttt{HistGradientBoostingClassifier} from the \texttt{scikit-learn}~\cite{scikit-learn} package as defined in \textcite{Finke:2023ltw}.
For this the default parameters set by \texttt{scikit-learn} are used except for the maximum number of iterations which is set to 200 to ensure that all classifiers reach a minimum in the validation loss.
The default parameters have a learning rate of 0.1, a maximum number of leaves of 31, no restriction on the maximum depth of the trees, and a maximum of 255 bins per feature.
The training of each classifier is repeated 50 times and averaged to ensure that the results are stable.

\subsection{Low level classifier}
The classifier trained on low level features is a transformer model as implemented in \textcite{drapes}.
The same model is used for both jets and is permutation invariant to the constituents of each jet and the order of the two jets.
The output of the encoder is summed to produce a single representation of the event, which is further processed by a feed forward network.
Four self attention layers are used with 256 hidden units and 2 heads.
Two cross attention layers are used with 256 hidden units and 2 heads.
The final feed forward network has 128 hidden units and has 2 hidden layers.
Models are trained for a maximum of 10 epochs with \textsc{AdamW}~\cite{loshchilov2019decoupledweightdecayregularization} as the optimizer with a learning rate of $10^{-4}$ and a weight decay of $10^{-5}$.
The learning rate is linearly ramped up from $10^{-6}$ to $10^{-4}$ over the first thousand batches.
A batch size of 256 is used.

\subsection{Pretraining}
The low level classifier is trained for 40 epochs on the full dataset with a batch size of 1024. 
In each dijet system half of the constituents are masked and the model is trained to predict the masked constituents following the set to set masked particle modelling paradigm~\cite{leigh2024tokenizationneededmaskedparticle}.

    \phantomsection
    \addcontentsline{toc}{chapter}{References}
    \printbibliography[title=References]

\end{document}